\documentclass[aps,pre,epsfig,showpacs,floats,twocolumn,amssymb,amsmath,floatfix,
groupedaddress,superscriptaddress]{revtex4-1}

\usepackage{graphicx} 
\usepackage{dcolumn} 
\usepackage{bm} 
\usepackage{float} 
\usepackage{bbm} 
\usepackage{color}
\usepackage{xcolor}
\usepackage{amsfonts}
\usepackage{lmodern} 
\usepackage{amsmath} 
\usepackage{amsmath,amssymb}

\begin{document} 
\title{
Self-generated gradients  stabilize the  hydrodynamic instabilities in active suspensions
}

\author{Mehrana R. Nejad} 
\affiliation{Department of Physics, Institute for Advanced Studies in Basic Sciences 
(IASBS), Zanjan 45137-66731, Iran} 

\author{Ali Najafi} 
\email[]{najafi@iasbs.ac.ir}
\affiliation{Department of Physics, Institute for Advanced Studies in Basic Sciences 
(IASBS), Zanjan 45137-66731, Iran} 
\affiliation{Research Center for Basic Sciences \& Modern Technologies (RBST), Institute for Advanced Studies in Basic Sciences, Zanjan, Iran}

\date{\today} 

\begin{abstract} 
Ordered phases emerged in active suspensions of polar swimmers are under long-wavelength 
hydrodynamic mediated instabilities.
In this letter, we show that chemical molecules  dissolved in aqueous suspensions, as an unavoidable part of most 
wet active systems, can mediate long-range interactions and subsequently stabilize the ordered phases. 
Chemoattractant  in living suspensions and dissolved molecules producing phoretic forces in synthesized Janus suspensions are reminiscent 
of such molecules.     
Communication between swimmers through the  gradients of such chemicals  generated by  individual swimmers,
is the foundation of  this stabilization mechanism. 
To classify the stable states of such active systems, we investigate the   detailed phase diagrams for two classes of systems with momentum conserving and non-conserving dynamics.  
Our linear stability analysis shows how  the stabilization mechanism  can work for swimmers with different dynamical properties, e.g.,  pushers or pullers and with various static 
characteristics, e.g., spherical, oblate or prolate geometries.   
\end{abstract} 
\maketitle

Understanding and explaining the physics of active matter have attracted many interests  recently \cite{vicsek2012collective,MarchettiRMP,RamaswamyAR,juelicher2007active}. 
Active  suspensions, both living and synthetic systems, are not bounded by equilibrium laws,  thus, show a variety of behaviors ranging from collective self organized motion (even in two dimension) \cite{simha,bickel2014polarization,vicsek1} to nontrivial rheological properties  \cite{bacteriasuperfluid,yeomans,rafaieffective,prost2015active,moradirheological}.  
Long-range rotational order, observed in active suspensions, are under strong dynamical instabilities mediated by hydrodynamic  interactions   
in low Reynolds wet systems  \cite{simha,ramaswamy-statphys}. This   instability is generic in the sense that, it is not affected by any short-range interaction but its underlying mechanism is  very sensitive to the hydrodynamic details of individual swimmers. 
For pushers (pullers),   bend (splay) fluctuations diverge and initiate the instability.
Interestingly and in contrast to this hydrodynamic mediated instability, there are   examples that  the ordered phases can be observed experimentally. 
 Furthermore, studying stabilization mechanisms  provides  guidelines for designing micro swimmers exhibiting collective ordered motion.
System-size dependent  fluctuations in elastic systems \cite{criticallength} and 2-D film confinement \cite{Joanny,leoni2010swimmers} provide  mechanisms that can stabilize the  instability. 


In this article, we show that chemical signaling between swimmers is another potential mechanism that can 
stabilize the  instabilities. Phenomena  of taxis (chemo, photo and {\it etc.})  that are vital activities in most living organisms \cite{adler1966chemotaxis} and 
phoretic interactions between active agents in  suspension of artificial  swimmers \cite{raminchemocollective,theurkauff2012dynamic} can be 
considered as examples of such chemical signaling.  
Depending on the details of system under investigation, chemotaxis itself can initiate Keller-Segel type instabilities \cite{kellersegelinstability,keller2}
but there are examples showing that 
phoretic interactions between active agents  can  lead to interesting collective behaviors  \cite{phoreticcollective1,raminchemocollective,catesphoreticcollective}.
Interplay between hydrodynamic and chemotaxis is investigated previously with a view on the effects of self generated flows  on the stability of isotropic phase of auto-chemotactic   swimmers \cite{lushi2012collective}. 
In a different regime and at the threshold of  hydrodynamic instabilities, we study the influence of chemotaxis on  long 
wave-length instabilities. 
 In both living and synthetic active matter, individual agents change their state of motion in response to a gradient in chemicals.  
Here we use a macroscopic phenomenological description in which, chemotaxis can be considered as  currents  proportional 
to the local gradient of concentration. Although, there are microscopic derivations that can support this picture \cite{anderson1,keller2}, chemotactic coefficients 
defined in this way can also be considered as parameters that can be measured experimentally. 
Extending the idea of phoretic Brownian particle \cite{stark} to both momentum conserving and non-conserving systems, we formulate a   continuum description for an active suspension and 
use linear stability analysis to determine the stability criteria.

Let us consider an interacting suspension  of swimmers moving in the presence of a concentration of chemical nutrients. 
Each swimmer is an axisymmetric particle with major and minor diameters given by $\ell$ and $\Delta\ell$  and it 
moves with an intrinsic speed given by $v_0$ along its major axis  denoted by a unit vector ${\bf m}$. In a mean field description,  
the dynamics of this suspension is described by single particle probability distribution function $\psi({\bf r},{\bf m},t)$, showing the 
probability to find a swimmer with orientation ${\bf m}$ in position ${\bf r}$ at time $t$. 
In addition to this distribution function, chemical concentration $c({\bf r},t)$ and velocity profile of the ambient fluid  ${\bf u}({\bf r},t)$, created by moving swimmers, are dynamical variables that need to be determined.  
The  
Smoluchowski equation for $\psi$ governs the dynamics of this interacting suspension:
\begin{equation}
\frac{\partial}{\partial t}\psi({\bf r},{ {\bf m}},t)+\nabla\cdot{\bf J}^t+
{\nabla}_m\cdot{\bf J}^m+({ {\bf m}}\times{\partial { {\bf m}}})\cdot{\bf J}^{m}_{\perp}=0,
\end{equation}
where $\nabla$ and $\nabla_m$ stand for positional and orientational gradients. 
Denoting by ${\bf D}$ and ${\bf D}_r$, the transnational diffusion tensor and rotational diffusion coefficient of the swimmers, the currents are given by:
\begin{eqnarray}
&&{\bf J}^t=[v { {\bf m}}+{\bf u}- D \nabla-(k_BT)^{-1} D \nabla U- \chi_t \nabla c]\psi,\nonumber\\
&&{\bf J}^m_{\bot}=-[D_r({ {\bf m}}\times{\partial { {\bf m}}})(1+\frac{U}{k_BT})+\chi_r{\bf m}\times\nabla c]\psi,\nonumber
\end{eqnarray}
and ${\bf J}^{m}=({\bf I}-{\bf m}{\bf m})\cdot({\bf \Omega}+A{\bf G})\cdot
{\bf m}\psi$. 
Chemotaxis transnational and rotational currents are  introduced through phenomenological 
 coefficients $\chi_t$ and $\chi_r$, both can have positive or negative values.
Symmetric and antisymmetric parts of the fluid velocity gradient are 
denoted by $2G=\nabla{\bf u}+(\nabla{\bf u})^T$ and $2\Omega=\nabla{\bf u}-(\nabla{\bf u})^T$ respectively. For axisymmetric swimmers we have $A=(1-\Delta^2)/(1+\Delta^2)$ \cite{jeffery}. 
In terms of  distribution function, we define density 
$\rho=\int d{\hat {\bf m}} \psi$, polarization 
${\bf P}=\int d{\hat {\bf m}} \psi{\hat {\bf m}}$ and nematic order parameter as:  
${\bf Q}=\int d{\hat {\bf m}} \psi({\hat {\bf m}}{\hat {\bf m}}-I/3)$. 
To take into account the hydrodynamic interactions, we   consider the dynamics of fluid.   
Denoting the viscosity of ambient fluid by $\eta$, fluid flow obeys Stokes equation $ \eta \nabla^2 \textbf{u} - \nabla \Pi = \nabla\cdot \sigma^a$, that is supplemented by 
incompressibility condition $\nabla \cdot {\bf u} =0$. Assuming that the swimmers are force-dipoles with strength $\zeta$, their contribution to the Stokes equations appears as an active stress $\sigma^a=\zeta{\bf Q}$ \cite{MarchettiRMP}.  For a dipolar swimmer,  we assign 
$\zeta=  6\pi\eta \ell^2 v_0\Delta_p$ where, $\Delta_p$ is a dimensionless number showing the strength of force-dipole associated with swimmer,   
for pusher $\Delta_p>0$ and for puller $\Delta_p<0$.  Alternatively, to account for  the hydrodynamic 
 effects,  an effective  two-body interaction between swimmers can also be considered \cite{farzin,yeomans3,behmadi}.
We also consider a  short-range  interaction potential as:  $U=-\frac{4}{3}\pi\ell^{3}U_0(1+\frac{\ell^{2}}{10}\nabla^2){\bf m} \cdot{\bf P}$ with $U_0>0$  to 
introduce   polar order in the suspension \cite{fazli}.

Considering both diffusion and convection, concentration of chemical molecules obeys the following equation:  
 $\partial_{t}c(\textbf{r},t) =-{\bf u}\cdot\nabla c+D_{c} \nabla ^{2} c - K(c) \rho(\textbf{r},t) +S$
where,  chemical molecules are injected to the medium through a uniform source term $S$ and swimmers act as  sinks for chemical molecules.   
  Reaction rate $K$,  is assumed to obey Michaelis-Menten kinetics, characteristic of catalytic reactions,   as:
$K(c)={ K_0c}{(c+c_M)^{-1}}$ where $K_0$ is the maximum reaction rate and $c_M$  denotes a concentration at which, the 
 reaction rate reaches to its half maximum value \cite{menten,MMExp1,MMExp2}. The kinetics we are considering here is the 
 simplest choice and there are other possibilities that we can consider as well without any crucial change in our final results.

To study the dynamics of fluctuations, we consider the case that our system fluctuates around a uniform distribution of both
chemical nutrients and swimmers and  set $c=c_0+\delta c$ and $\rho=\rho_0+\delta\rho$. 
For $t>1/D_c$, system reaches to steady state and denoting the   fluctuation
  wave vector by ${\bf q}$, the  chemical concentration and velocity profile can be obtained explicitly as:
\begin{equation}\label{2nd}
\tilde{{\bf u}} = \frac{ i \zeta }{ \eta q} 
\left(  {\hat {\bf q}} {\hat {\bf q}}   -{\bf I} \right)\cdot  \delta\tilde{{{\bf Q}}}\cdot{\hat {\bf q}},~~~
 \delta{\tilde c} =  \frac{-K(c_0) }{D_c q^2 +\rho_0 \partial_c K}\delta\tilde{\rho} ,
\end{equation}
where variables with tilde sign, show Fourier modes.  
In terms of chemical concentration, two different regimes of diffusion and reaction dominated can be distinguished.  
For sufficiently small concentrations $c_0\ll c_M$, the dynamics of chemical molecules is totally governed by reaction process as: 
$\delta{\tilde c} = -(c_0/\rho_0)\delta\tilde{\rho}$  but for larger concentrations ($c_0\ge c_M$), it is dominated by diffusion: $\delta{\tilde c} = -(K_0/D_cq^{2})\delta\tilde{\rho}$.
Having in hand the above results for $\delta{\tilde c}$ and 
${\tilde {\bf u}}$, we can eliminate them from the Smoluchowski equation and obtain  an equation that  governs the 
dynamics of $\psi$. Instead of $\psi$, we can study the dynamics of its moments $\rho$, 
${\bf P}$, ${\bf Q}$ and higher moments. Cutting the hierarchical equations obtained with this method at the 
second order moment,  will results in closed equations that show the dynamics of the system (details are given at the supplementary note).

\begin{figure}
\includegraphics[width=0.95\columnwidth]{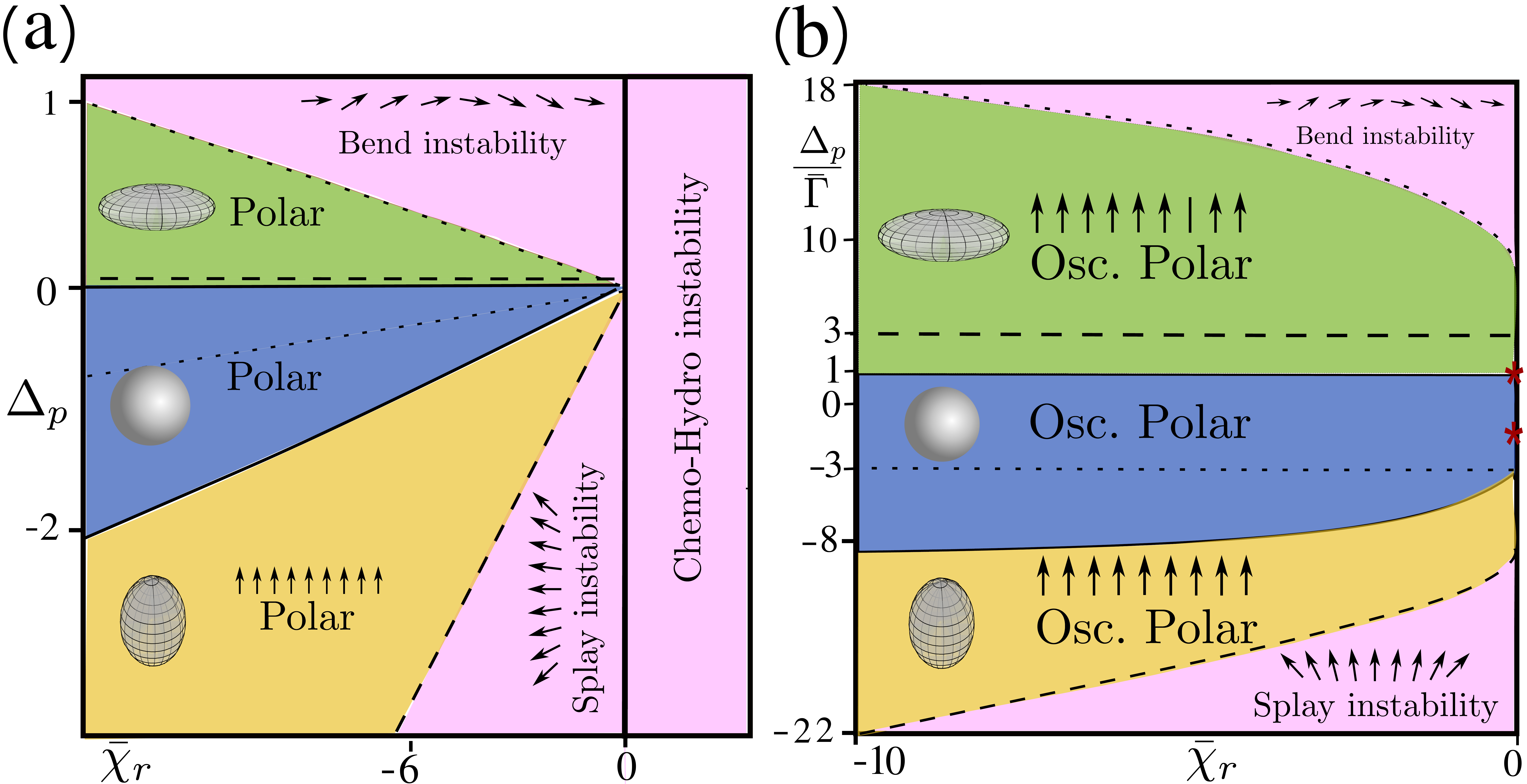} 
\caption{
(a) Stable regions for a $3$-D momentum conserving polar suspension. 
In terms  of  $\Delta_p$ and ${\bar \chi}_r$, stable regions are shown for rod-shape, spherical and disk-shape swimmers.  
Boundaries of the stable regions are shown by dotted, solid and dashed lines for rod-shape, spherical and disk-shape swimmers respectively.   
%
(b) Regions of stability for a momentum non-conserving  suspension interacting with substrate through friction coefficients ${\bar\Gamma}$
and $\Gamma^{\prime}<0$. On the vertical axes, two highlighted points specify
an interval $-{\bar\Gamma}_{min}<\Delta_p<{\bar\Gamma}_{max}$ where, the hydrodynamic
screening stabilizes the system for $\Gamma^{\prime}<0$ and ${\bar \chi}_t = {\bar \chi}_r = 0$.
With short-range chemotaxis-assisted  interaction, stable states will grow both for
positive and negative values of $\Gamma^{\prime}$. Here, we have set $E=-2$ and $Pe=1$.}
\label{fig1}
\end{figure} 

As a result of short-range interactions, a  transition from a homogeneous isotropic  state with 
$ \rho=\rho_0$ to a polar state with ${\bf P}=\rho_0 P_0{\hat {\bf n}}$  
 will take place. 
 %
To analyze  the stability of polar state, we add small fluctuations 
$\delta \textbf{n}$ and $\delta \rho$ to the steady state values then, study their linearized dynamics.
Introducing  the  Fourier transform for any  fluctuating field as $f({\bf  r},t)= \int{d{\bf q}}d\omega\tilde{f}({\bf q})e^{i({\bf q}\cdot{\bf r}-\omega t)}$ and denoting by $\theta$ the angle between ${\hat {\bf q}}$ and ${\hat {\bf n}}$, dimensionless modes can be obtained as: ${\bar \omega} = a_1 \ell q \cos{\theta} + i h_{\pm}$ where,  
\begin{equation}
\label{4th}
h_{\pm}=g(\theta)\!\bigg [1 \pm \! \bigg(\frac{\sin^2{\theta}({\bar \chi}_r\!\!+\!i a_2 \ell q \cos{\theta} )}{g(\theta)^2}+(1-\frac{2 {\bar \chi}_t }{g(\theta)})^2\!\bigg )^{\frac{1}{2}}\!\bigg ].
\end{equation}
Characteristic frequency $\omega_0=\frac{1}{2}\gamma\times(v_0/\ell)$ is used to make frequencies non-dimensional. Dimensionless phoretic  coefficients are defined by:
\begin{eqnarray}
&&{\bar \chi}_t= \frac{\chi_t}{2v_0\chi_0},~~{\bar \chi}_r= \frac{\chi_r}{\omega_0\chi_0},~~\chi_0=\frac{D_c\omega_0}{\rho_0 K_0 v_0 },~~\gamma=3\pi\rho_0\ell^3,
\nonumber\\
&&g(\theta)=\Delta_p \cos{2 \theta} + {\bar \chi}_t,~~~a_2/\Delta_p=2a_1=4\gamma^{-1}. \nonumber
\end{eqnarray}

%
For simplicity we first begin by considering the spherical swimmers $A=0$. Geometric effects for rod- or disk-like swimmers will be discussed at the end.
 The condition $\Im({\bar \omega})<0$ ($\Re({\bar \omega}) \neq 0$ or $=0$) shows the stability criterion  ( oscillating or non-oscillating)  and 
 the onset of instability is given by the condition $\Im({\bar \omega})>0$.  
 
 Well established results corresponding to the  
hydrodynamic mediated instability of  polar phase can be seen by  ignoring the chemotaxis in the above results \cite{simha,MarchettiRMP,PRLsaintillan2008}. 
Setting $ q = {\bar \chi}_t={\bar \chi}_r=0$, 
we see that $\Im({\bar \omega})=2 \Delta_p \cos{2 \theta}$,  showing that instability of pushers ($\Delta_p>0$) and pullers($\Delta_p<0$) are 
due to bend ($\theta=0$) and splay ($\theta=\pi/2$) fluctuations, respectively. 
 Chemotaxis mediated instabilities in the absence of hydrodynamic interactions can also be investigated by 
setting $\Delta_p=q=0$ that will  result in 
$\Im({\bar \omega}) = {\bar \chi}_t \big[ 1 \pm \Re (1+\frac{{\sin^2 \theta\bar \chi}_r}{{\bar \chi}_t^2})^{\frac{1}{2}}\big]$.  
Regarding the signs of ${\bar \chi}_r$ and ${\bar \chi}_t$, we can distinguish  different cases. When both of them are positive or one positive the other negative, it is easily seen that the homogeneous polar state is unstable. For ${\bar \chi}_t >0$, chemotaxis collapse occurs that eventually 
 makes the system inhomogeneous. For ${\bar \chi}_r > 0$, an instability in director (resulted from phoretic torque between swimmers) destroys polar order. Only the case where both chemotactic  coefficients are negative  
(${\bar \chi}_r<0,~{\bar \chi}_t<0$), stable polar state  is expected to observe in the system.  In this regime,  and for angles satisfying 
${\bar \chi_r}{\sin^2\theta}<-{{\bar \chi}_t}^2$, oscillating states can also be observed. Wave-number dependent oscillations of polar state in active matter  have been studied before \cite{PREbaskaran2008,PNASBaskaran2009}. Here, in addition to sound-like waves (terms proportional to $a_1$), wave-number independent ($q =0$) oscillations of polar state   is  observed.  
 
When both chemotaxis and hydrodynamics are considered, interesting  results will appear. 
Regarding the above discussion and in terms of chemotactic coefficients, 
we expect to see non-trivial results when ${\bar \chi}_r<0,~{\bar \chi}_t<0$. 
For  a fixed and  negative value of ${\bar {\chi}}_t$, fig.~\ref{fig1}(a) shows a phase diagram of possible phases that can appear in 
a non-confined  interacting suspension  at the limit of $q=0$.   Chemotactic coefficient ${\bar {\chi}}_r$ and strength of hydrodynamic interactions 
$\Delta_p$, are used to label the phase diagram.
As seen from the phase diagram, chemotaxis can not completely suppress hydrodynamic instabilities of pushers ($\Delta_p>0$) and 
a polar suspension of pushers is always unstable. Interestingly,  chemotaxis can suppress the splay fluctuations and 
stabilize the polar state for pullers ($\Delta_p<0$) and   
both static and oscillating  polar phases can be seen at the phase diagram. 
Oscillations of polarized state observed in this phase diagram are scale-free in a sense that their frequency do not depend on 
wave-vector. 

\begin{figure}
\includegraphics[width=0.95\columnwidth]{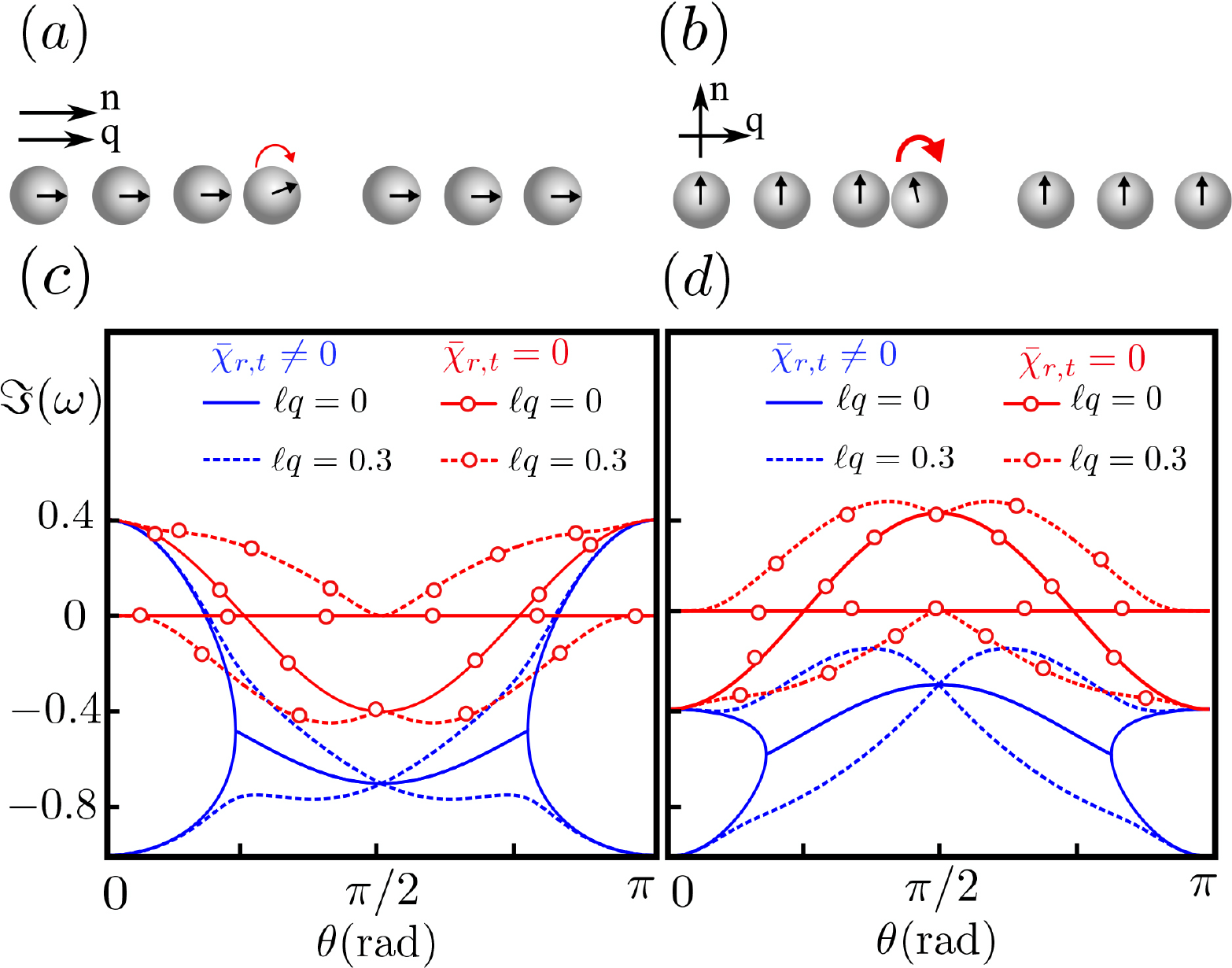} 
\caption{(a) and (b) demonstrate how  chemotaxis  tends to suppress both bend and splay distortions in a suspension of spherical swimmers. As we can see,   
restoring torque (shown by curved red arrow) is  more stronger for the case of  splay fluctuations. 
(c) and (d)  show the growth rate, $\Im(\omega)$, as a function of  wave angle $\theta$ in a suspension of nearly aligned pushers and pullers, respectively. 
In the absence of chemotaxis  (circled-red), bend distortions ($\theta=0,\pi$) make pushers suspension unstable but for pullers it is splay fluctuation 
($\theta=\pi/2$) that initiate  the instability.     
Effects due to chemotaxis, shown as   blue lines,  strongly (weakly) diminish splay (bend) fluctuations  for infinite and finite systems. 
Numerical values are  $a_2 = 8,~|\Delta_p| = 0.2,~{\bar \chi}_r=-0.6$ and ${\bar \chi}_t = -0.5$. }
\label{fig2}
\end{figure}

To have an intuitive picture about the stabilization mechanism,  
figures \ref{fig2}(a) and (b) show a collection of nearly parallel swimmers that are under small 
bend and splay fluctuations, respectively. As seen from 
figures and as a result of such director distortions, density fluctuations will appear in the system.  In terms of director fluctuations $\delta n$, density 
fluctuation for the case of splay is a first order effect but it is  second order for bend distortion. 
In both  cases,  considering  density fluctuations shows that, reoriented swimmers are much 
affected by the  swimmers from left side  where, their overall  chemotactic torque tends to diminish  fluctuations. Fluctuations suppression is much stronger for the splay case. To obtain this result, 
we have used the relation $-{\bar \chi_r}{\bf m}\times\nabla c$ with ${\bar \chi_r}<0$ for chemotactic angular velocity and have assumed that  at steady state, each  swimmer consumes  chemical molecules $(K_0>0)$ and  produce a radial gradient in chemicals. 
To prevent chemotactic collapse, it is necessary to consider ${\bar \chi}_t<0$. 
For the case of bend fluctuations and at the first order of $\delta n$, 
chemotactic torques acting on distorted swimmers from left and right sides will 
cancel each other. Considering the higher order corrections, chemotaxis  
tends to diminish the fluctuations but,   it is not so strong  to  remove the instability mediated from bend fluctuations in a system of pushers.  

To investigate the stability of polar state in  a  system with finite size, we have plotted in fig.~\ref{fig2}(c) and (d), 
the  growth rate $\Im(\omega)$ as a 
function of  $\theta$ for different values of $q$.  
Regarding the instability criterion $\Im(\omega)>0$,   fig.~\ref{fig2}(c) 
shows that for pushers and at the absence of chemotaxis, the instability comes from bend modes ($\theta=0,\pi$). 
Here, chemotaxis can suppress the fluctuations and make the 
instability angles more narrower but it is not able to totally remove the instability.
This conclusion is valid for both infinite and finite systems.
Fig.~\ref{fig2}(d) shows that for pullers, and at ${\bar \chi}_r={\bar \chi}_t=0$, splay modes ($\theta=\pi/2$) diverges and 
initiate hydrodynamic instability. In this case chemotaxis 
  can suppress the splay fluctuations  for both infinite 
and finite system of pullers and eventually stabilize the system.  Note that  parameters are chosen from the stable region of phase diagram.  

Elasticity that is identified by dimensionless  bend and splay moduli ${\bar K}_b$  and ${\bar K}_s$,  is  another interesting effect that can suppress the fluctuations.   Including elasticity in our model, function 
$g(\theta)$ should be replaced by  $g(\theta)=\Delta_p \cos{2 \theta} + {\bar \chi}_t - (q \ell)^2 ({\bar K_s} \sin^2 \theta + {\bar K_b} \cos^2 \theta)$ in equation 3.
As shown in reference \cite{criticallength}, elasticity introduces a length 
$L_b=\ell\sqrt{{{\bar K}_b}/{\Delta_p}}$, that systems with smaller sizes $L<L_b$, are stable against hydrodynamic fluctuations.
Presence of chemotaxis does not change this picture for a suspension of pushers, but it
 enhances  the threshold length scale. 
This elasticity induced stability mechanism for finite systems, works for the unstable part  of the  phase diagram 
presented in fig.~\ref{fig1}(a).
 
To investigate the stability of isotropic  state, we  study the dynamics of fluctuations around a steady state given by 
 $P=Q=0$.
For diffusion-dominated regime, linearizing the 
dynamical equations leads to the following modes for the fluctuations at $q=0$:
\begin{equation}\label{5th} 
{\bar{\omega}}\! ={\bigg \{ \! i{\bar \chi}_t -{i \bar{D}} \pm i \big(\frac{2}{3}{\bar \chi}_r +({\bar \chi}_t+{\bar{D}} )^2\big)^{\frac{1}{2}}},
{\frac{4i}{5} A \: \Delta_p -{6 i {\bar D}_0}\bigg \}}, \nonumber
\end{equation}
where ${\bar D}_0=D_r/\omega_0$, $\bar{D} \!= \! {\bar D}_0 (1-{\rho_0}/{\rho^*})$ and $\rho^*=9/(4\pi U_0\ell^3)$. 
Stability of an isotropic suspension of swimmers depends not only on their type (pusher or puller), but also on their shape through parameter $A$.
In the absence of chemotaxis and for $A \Delta_p > {15 {\bar D}_0}/{2}$, the isotropic state is unstable both for a suspension of  pushers 
($\Delta_p >0$) with $A >0 $ and for  pullers $(\Delta_p <0)$ with $A <0$  \cite{PRLsaintillan2008,PNASBaskaran2009}. 
Isotropic suspension of spherical swimmers ($A = 0$) is always stable. 
As one can see from the above equation,  modes associated to hydrodynamic and chemotaxis are independent and chemotaxis is not able to remove the  
instabilities . Taking into account the stability criterion for both hydrodynamic and chemotaxis part, we see that for 
$A \Delta_p <{15 {\bar D}_0}/{2}$, stable isotropic state 
can be observed under the condition ${\bar \chi}_t < \bar{D}$ and ${\bar \chi}_r<-6  \bar{D} {\bar \chi}_t$.

Extending all the above results for reaction-dominated  regime, shows that 
 chemotaxis does not have any strong effect on the phase portrait of  both  polar and isotropic  momentum conseving   suspensions. 
In this regime,  any local decrease in chemical molecules  does not 
have enough time to diffuse and propagate to the position of other swimmers and subsequently chemotaxis  
 is not able to remove hydrodynamic instabilities.

 Friction with a substrate is another interesting and 
 important factor in many experiments. Such friction  can remove the instabilities by screening the long-range hydrodynamic interactions \cite{amin}. 
To study the dynamics of    a suspension that is in contact with substrate,  we replace the Stokes  equation by:  
$-\Gamma{\bf u}-\nabla\Pi=\zeta_{2D}\nabla\cdot {\bf Q}+\Gamma'{\bf P}$ where $\Gamma$($>0$) and $\Gamma'$ are two phenomenological friction
coefficients for the  fluid and swimmers respectively \cite{friction1,friction2}.  For diffusion-dominated regime, neglecting the effects of convection,  as a result of hydrodynamic screening,  
polar suspension is always stable. For reaction-dominated regime both hydrodynamic and chemotaxis appear as short-range effects. Stable states of 
the suspension for this case, is  presented in  phase-diagram   fig.~\ref{fig1}(b) where, we investigate the stability criterion for different values of friction and rotational chemotaxis coefficients. Friction coefficient enters through a 
dimensionless variable given by $\Delta_p/{\bar \Gamma}$ with ${\bar \Gamma}=\Gamma\ell^2/\eta$.   At the absence of chemotaxis (${\bar \chi}_r={\bar \chi}_t=0$) and for 
$\Gamma'<0$,  friction can stabilize the suspension. This stability occurs in an interval given by 
$-{\bar \Gamma}_{min}<\Delta_p<{\bar \Gamma}_{max}$, denoted by  two highlighted points on the vertical axis.  By turning on the chemotaxis, 
both rotational and translational, available stable states will grow  for $\Gamma'<0$.  For $\Gamma'>0$ and at the presence of chemotaxis, 
there is also a stable region that is shown in fig.~\ref{fig1}(b). 

To analyze the stability criteria for non-symmetric swimmers, we set $A \neq 0$ and study the fluctuations
spectrum.  Detail analysis show that, in comparison to
the spherical swimmers, hydrodynamic fluctuations are weaker in both rod-shape pullers and disk-shape pushers. This result holds for both momentum conserving and momentum non-conserving systems. As a result, depending on the  geometry of swimmers, chemotaxis can stabilize both pusher and puller suspensions. Results for both prolate and oblate swimmers are reflected in fig.~\ref{fig1}.

To estimate the   range  of   chemotactic 
coefficients in micron scale systems, we note that  the chemotactic velocity has the same order of magnitude as the swimming speed thus,  $\chi_t\sim v_0\ell/c$. For a swimmer with $v_0=50\mu{\text m}/{\text s}$ and $\ell=5\mu{\text m}$, moving 
in a $10{\mu}{\text M}$ concentration of food molecules with $D_c\sim5\times 10^{-10}{\text m}^2/{\text s}$, 
we can estimate the dimensionless chemotactic coefficients defined  in equation (3) as: ${\bar \chi}_t\sim{\bar \chi}_r\sim {\cal O}(1)$. 
Here we have used  $K_0\sim 10^3{\text s}^{-1}$ and $\rho_0\sim 10^{16}{\text m}^{-3}$. This estimation shows that our choice of 
parameters in fig.~\ref{fig1}, can cover most real systems.


In conclusion, we have studied the role of chemotactic interaction in both wet and dry active systems and have shown that for
both pushers and pullers, chemotaxis can suppress fluctuations. In a bulk of fluid, this suppression is much stronger for pullers and
can develop a stable region in their phase diagram for various geometries of swimmers. For pushers in bulk fluid, chemotaxis can stabilize suspension of disk-shape swimmers. In the presence of a substrate and for small chemical P\'eclet number, long-range chemotactic interaction can stabilize both puller and pusher suspensions when ${\bar \chi}_t$,${\bar \chi}_r<0$. In the case of finite P\'eclet number (presented in fig.~\ref{fig1}(b),  chemotaxis-assisted interaction can stabilize hydrodynamic fluctuations.

Helpful discussions with S. Ramaswamy are gratefully acknowledged.


%
\end{document}